\documentclass[final,3p,times,twocolumn]{elsarticle}
 \biboptions{comma,sort&compress}
\usepackage{here}
 \usepackage{graphicx}
  \usepackage{epsfig}

\usepackage{amsmath}

\def\nin{\noindent}
\def\beq{\begin{equation}}
\def\eeq{\end{equation}}
\def\bea{\begin{eqnarray}}
\def\eea{\end{eqnarray}}
\def\nnb{\nonumber}

\journal{Nuc. Phys. (Proc. Suppl.)}

\begin{document}

\begin{frontmatter}



\title{Towards a Numerical Implementation of the Loop-Tree Duality Method}

 \author[label1]{Sebastian Buchta\corref{cor1}}
  \address[label1]{Instituto de F\'{\i}sica Corpuscular, Universitat de Val\`{e}ncia 
- Consejo Superior de Investigaciones
Cient\'{\i}ficas, 
\\
Parc Cient\'{\i}fic, E-46980 Paterna, Valencia, Spain}
\cortext[cor1]{Speaker}
\ead{sebastian.buchta@ific.uv.es}

 \author[label1]{Grigorios Chachamis}
\ead{grigorios.chachamis@ific.uv.es}

 \author[label2]{Petros Draggiotis}
  \address[label2]{Institute of Nuclear and Particle Physics, NCSR ``Demokritos'', Agia Paraskevi, 15310, Greece}
\ead{petros.draggiotis@gmail.com}

 \author[label1]{Ioannis Malamos}
\ead{ioannis.malamos@ific.uv.es}

 \author[label1]{Germ\'{a}n Rodrigo}
\ead{german.rodrigo@csic.uv.es}


\begin{abstract}
\noindent
We review the recent progress on the numerical implementation of the Loop-Tree Duality Method (LTDM) for the calculation of scattering amplitudes. A central point is the analysis of the singularities of the integrand. In the framework of the LTDM some of these singularities cancel out. The ones left over are dealt with by contour deformation. We present details on how to achieve this as well as first results.

\end{abstract}


\end{frontmatter}


\section{Introduction}
\nin
When calculating NLO (NNLO) cross-sections one needs to consider the tree- and loop-contributions separately. Especially loops with many external legs prove to be challenging. Considerable progress has already been made in order to attack this problem: OPP- Method, Unitarity Methods, Mellin-Barnes Representation, Sector Decomposition \cite{Gehrmann:2010rj}.  The advantage of these methods is that they made possible what was impossible before, but still a lot of effort has to be put in to cancel infrared singularities among real and virtual corrections. Additional difficulties arise from threshold singularities that lead to numerical instabilities. 
The Loop-Tree Duality method aims towards a combined treatment of tree- and loop- contributions. Therefore the Loop-Tree Duality method casts the virtual corrections in a form that closely resembles the real ones.
\section{Loop-Tree Duality at one loop}
\nin
The most general, dimensionally regularized one-loop scalar integral can be written as \cite{Catani:2008xa}:
\beq
L^{(1)}(p_1, p_2,\dots , p_N) = \int\limits_{\ell_1}\prod\limits_{i=1}^NG_F(q_i)
\label{eq:olamp}
 \eeq
  with the Feynman propagator $G_F(q_i) = [q_i^2-m_i^2+i0]^{-1}$, internal momenta $q_i = \ell_1 + p_1 + \dots + p_i = \ell_1 + k_i$ and shorthand integral notation $\int_{\ell_1} = -i\int d^d\ell_1/(2 \pi)^d$. As a first step, one performs the integration over the complex energy components of the loop four-momentum by applying the residue theorem. The integration contour is chosen such that it encloses the poles with positive energy and negative imaginary part, see figure below:
  \begin{figure}[H]
  \centering
  \includegraphics[scale=1]{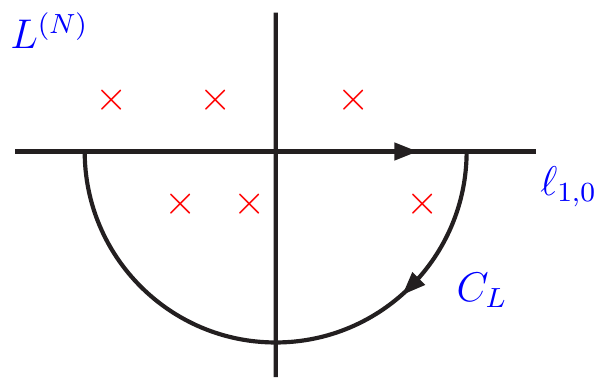}
  \caption{Location of poles and integration contour $C_L$ in the complex $\ell_{1,0}$-plane.}
    \label{fig:contour}
  \end{figure}
The residue theorem is employed by taking the residues of the poles inside of the contour and summing over them. Given an appropriate gauge choice the integrand in eq. (\ref{eq:olamp}) contains only simple poles. Thus the residue of an individual pole is done by taking the residue of a single propagator and evaluating the other propagators at the position of the residue
\beq
\text{Res}_{\text{Im}\{q_{i,0}\}<0}\frac{1}{q_i^2-m_i^2+i0}=\int d\ell_1\delta_+(q_i^2-m_i^2)\label{eq:res}
\eeq
\bea
\left.\prod\limits_{j \neq i}G_F(q_j)\right|_{\text{i-th pole}}=\prod\limits_{j \neq i}\frac{1}{q_j^2-m_j^2-i0\eta(q_j-q_i)}\label{eq:GD}\nnb\\
\equiv \prod\limits_{j \neq i}G_D(q_i; q_j).
\eea
The subscript ``+'' on the right hand side of eq. (\ref{eq:res}) indicates that the positive-energy solution is to be taken. Furthermore $\eta$ is a future-like vector, i.e. $\eta^2\geq 0$, $\eta_0 > 0$. It is dependent on the choice of coordinate system, however it cancels out once one adds all dual contributions. Hence physical objects like scattering cross sections will stay frame-indepedent. Evaluating the ``non-cut'' propagators at the position of the pole leads to a modification of the usual Feynman prescription. In eq. (\ref{eq:GD}) it is shown that instead one ends up with the so called ``dual prescription'' which serves to keep track of the correct sign of the i0-prescription of the corresponding propagator. Collecting all the pieces and putting them together, one arrives at
\beq
L^{(1)}(p_1,p_2,\dots ,p_N)=-\sum\int_{\ell_1}\tilde{\delta}(q_i)\prod\limits_{\substack{j=1\\ j\neq i}}^NG_D(q_i;q_j)
\eeq
with $\tilde{\delta}(q_i)=2\pi i\delta_+(q_i^2-m_i^2)$. Thus, by virtue of employing the residue theorem, it is possible to rewrite a one-loop amplitude as a sum of single-cut phase-space integrals over the loop-three-momentum. The i-th dual contribution has the i-th propagator set on-shell while the left over Feynman propagators get promoted to Dual propagators.
\beq
L^{(1)}(p_1,p_2,\dots ,p_N)=\nnb
\eeq
  \begin{figure}[H]
  \centering
  \includegraphics[scale=0.8]{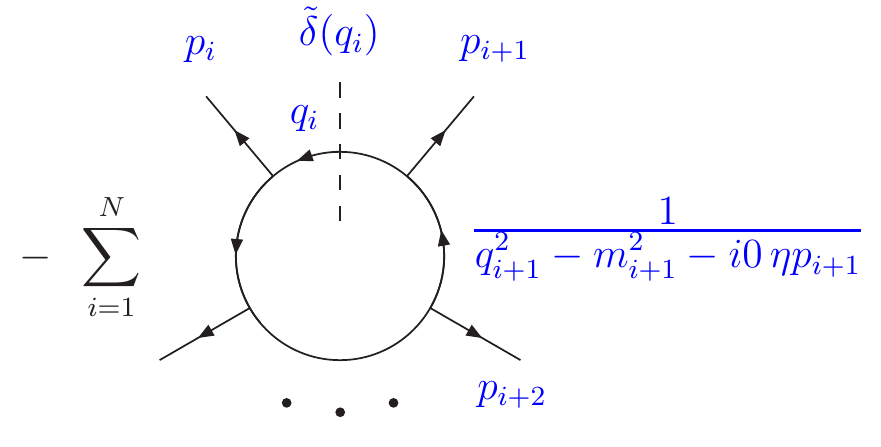}
  \caption{Graphical representation of the solution of the LTDM at one-loop.}
  \end{figure}
  The LTDM features a couple of interesting properties: 
  \begin{itemize}
  \item Number of single cut Dual Contributions equals the number of legs, this way a loop diagram is fully opened to tree diagrams. 
  \item The singularities of the loop diagram appear as singularities of the Dual Integrals. 
  \item Tensor loop integrals and physical scattering amplitudes are treated in the same way since the Loop-Tree Duality works only on propagators. 
  \item Virtual corrections are recast in a form, that closely parallels the contribution of real corrections.
  \end{itemize}
This is the formalism for the one-loop case. Solutions for more complicated situations like multiple loops \cite{Bierenbaum:2010cy} or higher order poles \cite{Bierenbaum:2012th} are described in the respective references.

\nin
\section{Singular behavior of the loop integrand}
\begin{figure}[H]
  \centering
  \includegraphics[scale=0.5]{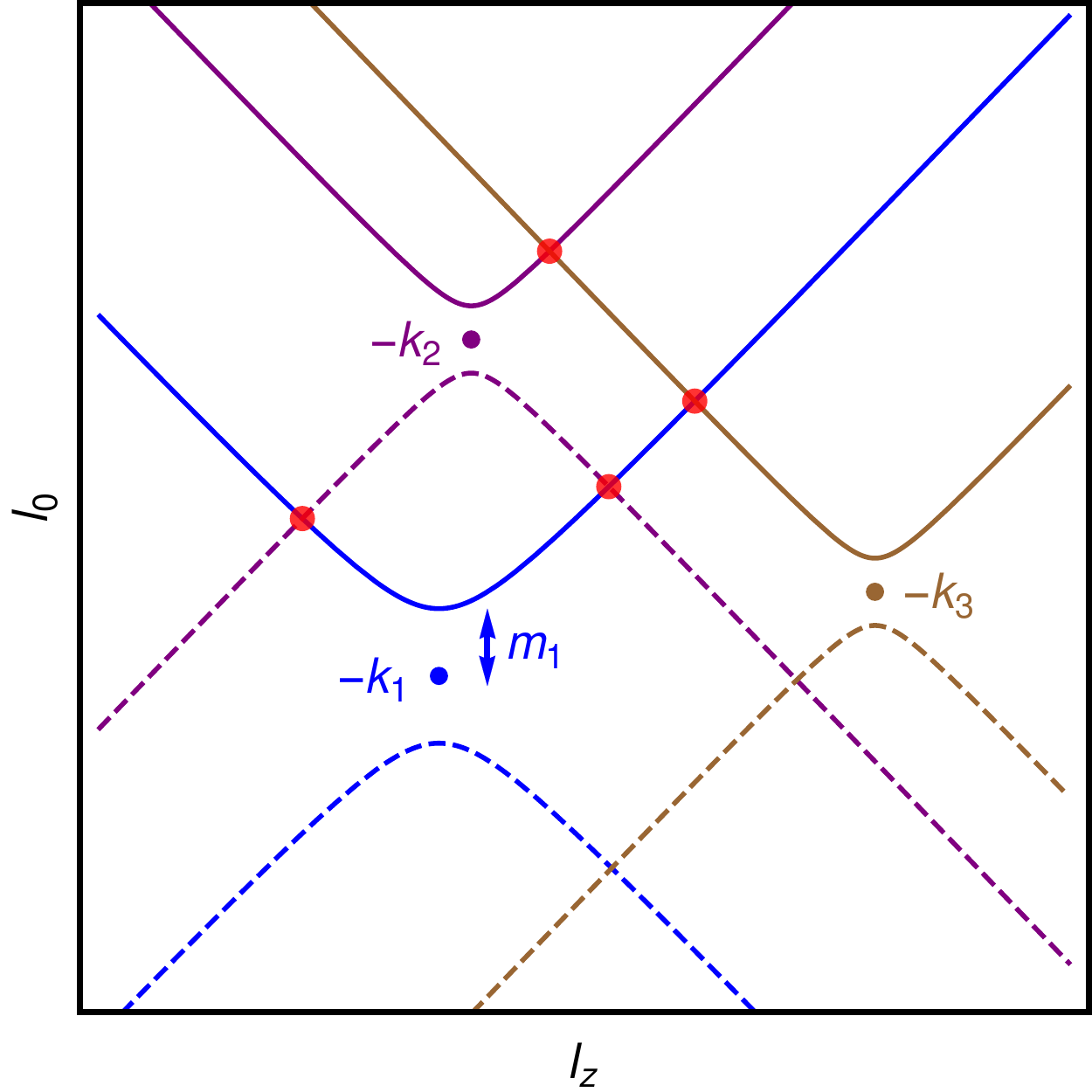}
  \caption{On-shell hyperboloids for three arbitrary propagators in Cartesian coordinates.}
    \label{fig:lightcones}
  \end{figure}

\nin
As a preparatory step it will prove useful to introduce an alternative way of denoting the dual propagator. This will give a more natural access to its singularities.
\beq
\tilde{\delta}(q_i)G_D(q_i;q_j)=2\pi i\frac{\delta(q_{i,0}-q_{i,0}^{(+)})}{2q_{i,0}^{(+)}}\frac{1}{(q_{i,0}^{(+)}+k_{ji,0})^2-(q_{j,0}^{(+)})^2}
\label{eq:rwprop}
\eeq
with $k_{ji}=q_j-q_i$ and $q_{i,0}^{(+)}=\sqrt{\mathbf{q_i}^2+m_i^2-i0}$.\\
In fig. (\ref{fig:lightcones}), the on-shell hyperboloids of three propagators in loop-mometum-space are sketched.
The loop integrand becomes singular at hyperboloids with $q_{i,0}^{(+)}=\sqrt{\mathbf{q_i}^2+m_i^2-i0}$ (solid lines) and $q_{i,0}^{(-)}=-\sqrt{\mathbf{q_i}^2+m_i^2-i0}$ (dashed lines) and origin in $-k_{i,\mu}$. Applying the LTDM is equivalent to integrating along the \textit{forward hyperboloids}. The intersection of two forward hyperboloids (solid lines) leads to singularities that will cancel among dual integrals. The intersection of a forward with a backward hyperboloid (solid line with dashed line) leads to a singularity that remains and therefore has to be dealt with by contour deformation.\\
Singularities appear where the denominator of the dual propagator becomes $0$. Since the denominator in eq. (\ref{eq:rwprop}) has been rewritten as a difference of squares, one can use the third binomic formula to take it apart and extract the conditions for which the singularities show up:
\bea
q_{i,0}^{(+)}+q_{j,0}^{(+)}+k_{ji,0}=0\label{eq:condel}\\
q_{i,0}^{(+)}-q_{j,0}^{(+)}+k_{ji,0}=0\label{eq:condhyp}
\eea
Eq. (\ref{eq:condel}) describes an ellipsoid in the loop three-momentum and demands $k_{ji,0}<0$. An ellipsoid is the result of the intersection of a forward with a backward hyperboloid. The origins of the hyperboloids are separated in a time-like fashion, expressed by the conditions
\beq
k_{ji}^2-(m_j+m_i)^2\geq 0, \quad k_{ji,0}<0.
\eeq
The singularity described by eq. (\ref{eq:condhyp}) has a hyperboloid shape as a result of the intersection of two forward on-shell hyperboloids of space-like separation. The corresponding condition is
\beq
k_{ji,0}^2-(m_j-m_i)^2\leq 0.
\eeq
Here, $k_{ji,0}$ may be positive or negative.
\nin
\section{Numerical Implementation}
\nin
The singularities associated with ellipsoid intersections require contour deformation. The following one-dimensional example illustrates how this can be achieved. The function
\beq
f(\ell_x)=\frac{1}{\ell_x^2-E^2+i0}
\eeq
has poles at $\pm (E-i0)$. Simply integrating along the real axis would lead to infinities. Therefore the integration contour has to be deformed to go around the poles. A suitable contour deformation would be:
\beq
\ell_x\rightarrow\ell_x'=\ell_x+i\lambda\ell_x\exp\left(-\frac{\ell_x^2-E^2}{2E^2}\right)
\label{eq:cdef2d}
\eeq
The parameter $\lambda$ serves to scale the deformation along the imaginary axis. At the position of the pole the exponent becomes 0 and thus the exponential function hits its maximum, which is $1$. Far away from the poles, the exponent is a large negative number, hence exponentiating it suppresses the deformation.
  \begin{figure}[H]
  \centering
  \includegraphics[scale=0.6]{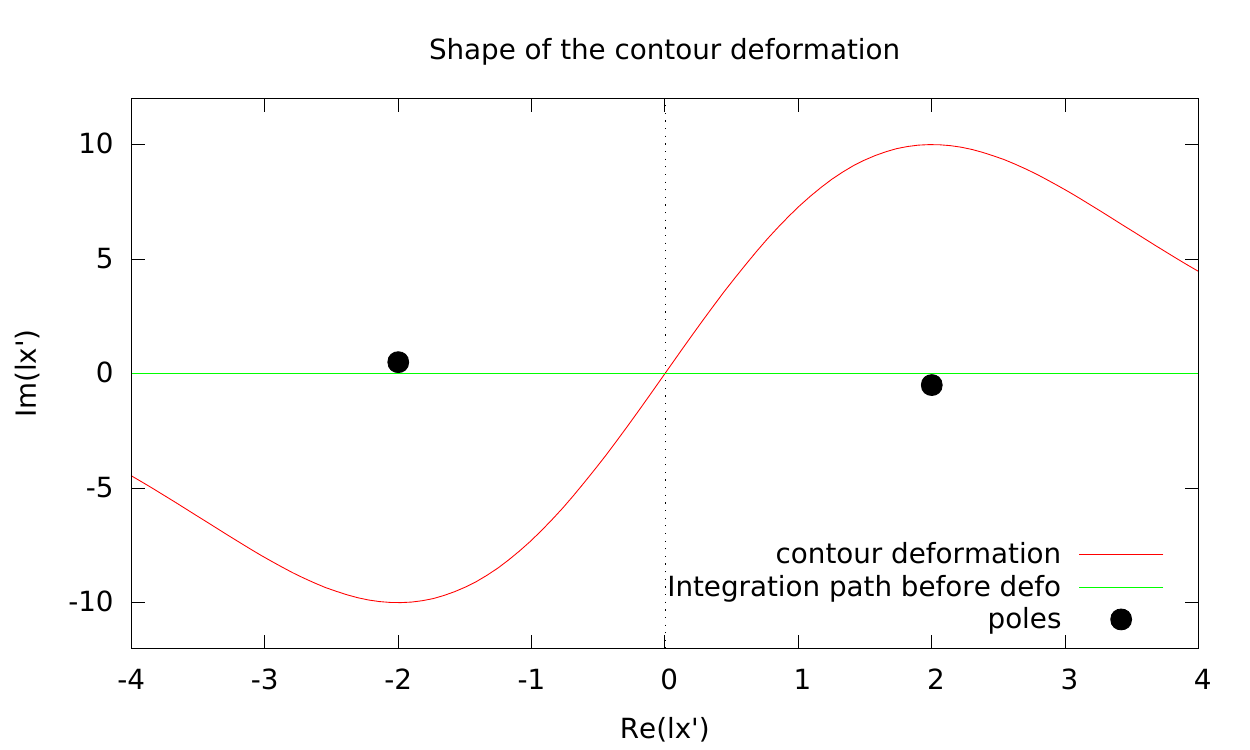}
  \caption{Contour deformation as in eq. (\ref{eq:cdef2d}) for $E=2$ and $\lambda=5$.}
    \label{fig:Defo2D}
  \end{figure}
 For the integration in loop three-momentum-space eq. (\ref{eq:cdef2d}) needs to be generalized to three dimensions. This is done by modifying the exponent and, of course, promoting $\ell_x$ to $\vec{\ell}$:
  \beq
  \vec{\ell}\rightarrow\vec{\ell}+i\lambda\vec{\ell}\exp\left(-\frac{G_D^{-2}}{\text{width}}\right)
  \eeq
  $G_D^{-2}$ now plays the same role as $\ell_x^2-E^2$ did in eq. (\ref{eq:cdef2d}). The parameter ``width'' allows to control the width of the deformation, thus giving additional control over it. 
    \begin{figure}[H]
  \centering
\includegraphics[scale=0.52]{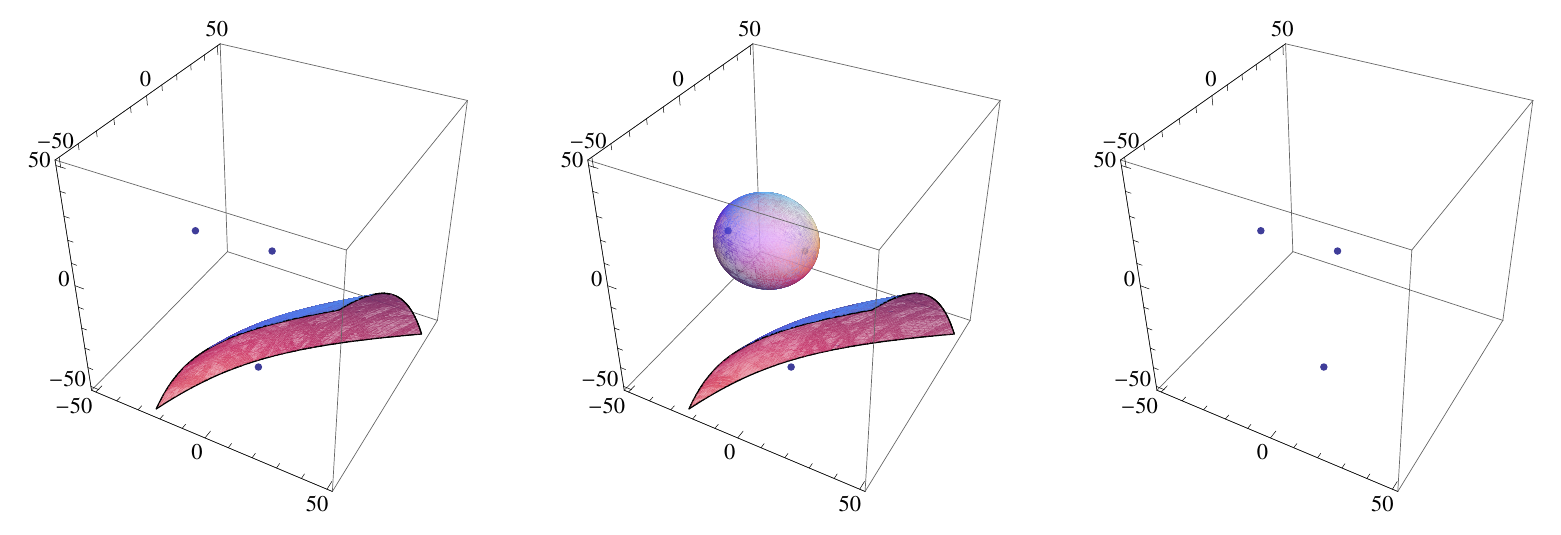}
  \caption{Hyperboloid and ellipsoid singularities in a triangle.}
    \label{fig:dualcont}
  \end{figure}
  The example in fig. (\ref{fig:dualcont}) shows the location of the singularities as they might appear for a triangle. Every box corresponds to a Dual contribution, i.e. cut. The easiest case is the third box, which doesn't display any singularities of any type and thus can be integrated straightforwardly. In the first contribution there is only a hyperboloid singularity, which cancels with the hyperboloid singularity from the second contribution. But since the second contribution features an ellipsoid singularity as well, this contribution requires contour deformation. In turn, in order to \textit{preserve the cancellation} of the hyperboloid singularities, contribution one will need the exact same deformation, \textit{despite} no ellipsoid being present there. This means that hyperboloid singularities act as a coupling between different dual contributions. The coupled contributions need to be deformed according to all ellipsoid singularities which they \textit{both} have.\\
  In the massless case collinear singularities arising from forward-forward intersections cancel among Dual integrals, because the on-shell hyperboloids of fig. (\ref{fig:lightcones}) would be tangential there. Nonetheless, collinear and soft singularities from forward-backward intersections remain, but they are restricted to a finite region that can be mapped to the real phase-space emission \cite{Buchta:2014dfa}.
  \nin
\section{Results and Conclusion}
\begin{figure}[H]
  \centering
\includegraphics[scale=0.67]{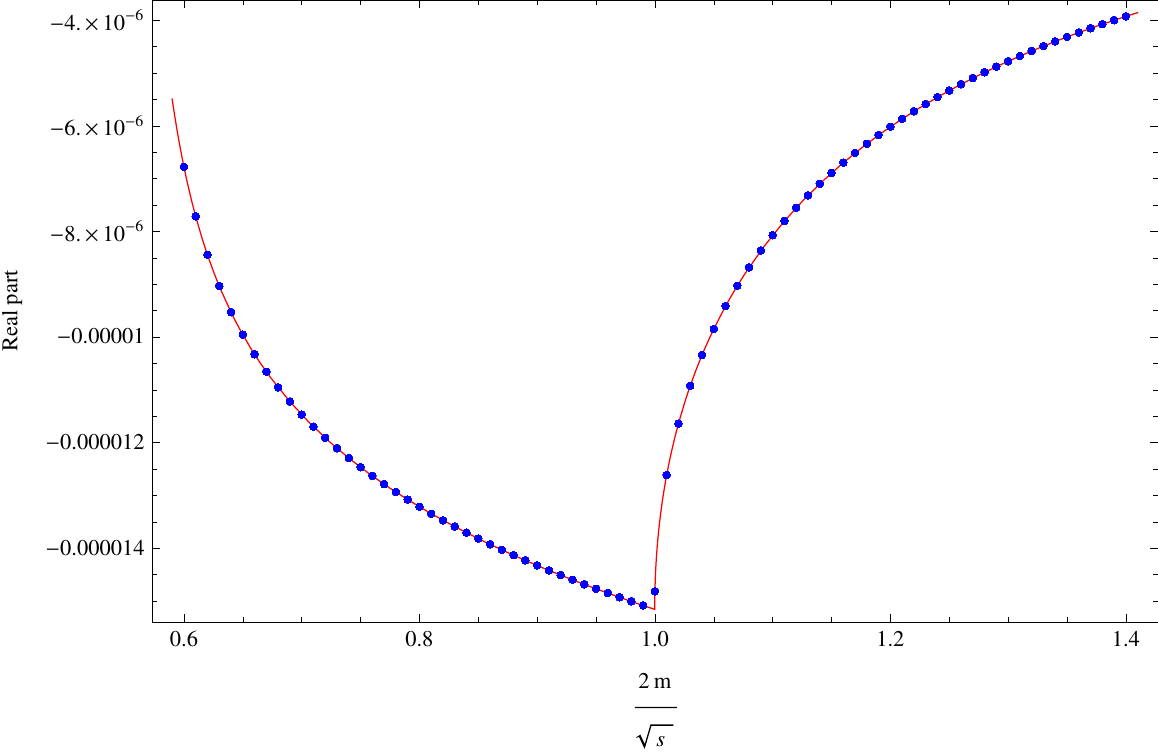}
  \caption{Real part. The curve is LoopTools, the dots are the LTDM.}
    \label{fig:thre}
  \end{figure}

\begin{figure}[H]
  \centering
\includegraphics[scale=0.67]{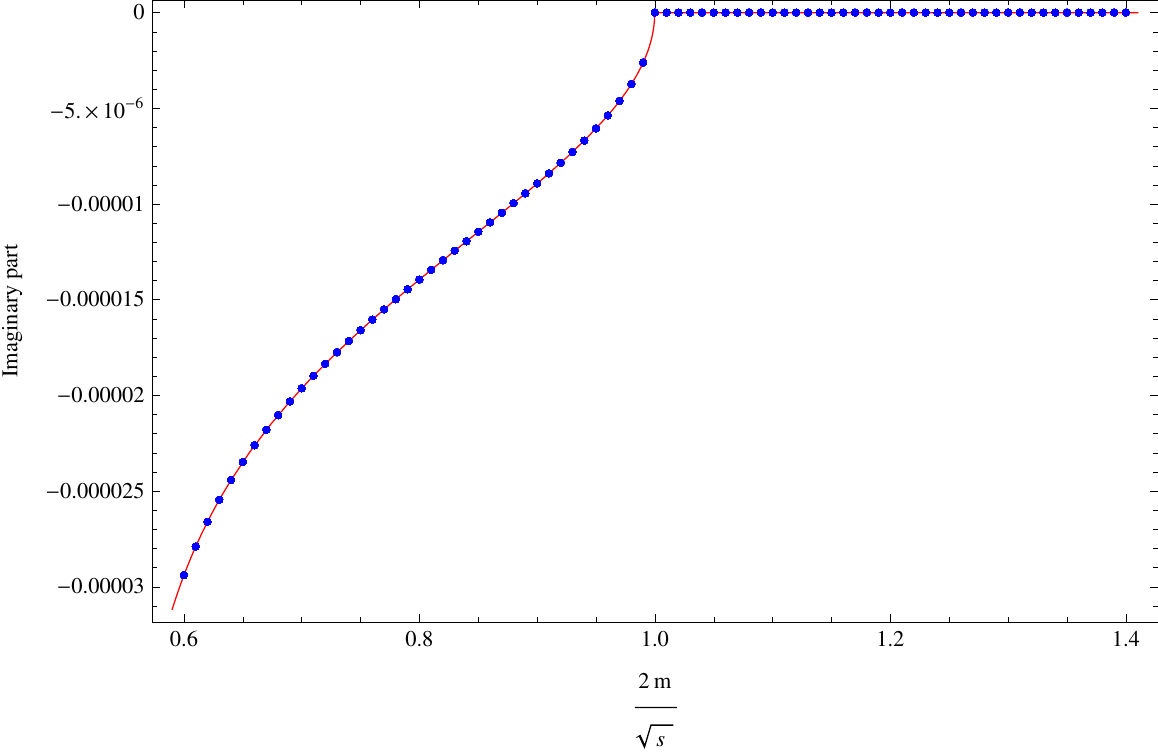}
  \caption{Imaginary part. The curve is LoopTools, the dots are the LTDM.}
    \label{fig:thrim}
  \end{figure}

\nin
A computer program has been written in C++. The numerical integration step is performed with Cuhre from the Cuba-Library \cite{Hahn:2004fe}. Analytic values for comparison are generated with LoopTools (LT) \cite{Hahn:1999wr}. The code runs on an Intel i7 desktop machine with 3.4GHz. It is capable of calculating triangles, boxes and pentagons with no deformation needed with 4 digits precision in 0.5s. The table below shows an explicit result where a pentagon with deformation has been calculated with 4 digits precision in 25s time.
\begin{table}[H]
\setlength{\tabcolsep}{0.3pc}
    {\small
\begin{tabular}{lllll}
&\\
\hline
 &Real part & Re Err. & Imaginary part & Im Err.  \\
\hline
LT & -1.001066E-10 & 0 & -5.208136E-10 & 0\\
LTDM & -1.001089E-10 & 9.1E-16 & -5.208556E-10 & 9.1E-16\\
\hline
\end{tabular}
}
\label{tab:param}
\end{table}
Another test the program has been put through is a scan of the region around threshold. To produce the two graphs in fig (\ref{fig:thre}) and (\ref{fig:thrim}), a triangle has been taken; the center-of-mass energy $s$ has been kept constant while the mass $m$ (which was taken to be identical for all three internal lines) was varied.
  From the plots one can see that LoopTools' analytical values are well matched by the computer program, this indicates good precision. There is no loss of precision or increase in calculation time close to threshold.\\
  Despite still being in development, the program already shows competitive speed and precision when calculating loop integrals. At the moment, only diagrams up to pentagons have been checked, but within the framework of the LTDM, extension to graphs with more external legs is straightforward and easy to realize.
\section*{Acknowledgements}
\nin
This work has been supported by the Research Executive Agency (REA)
of the European Union under the Grant Agreement number PITN-
GA-2010-264564 (LHCPhenoNet), by
the Spanish Government and EU ERDF funds (grants FPA2011-23778 and CSD2007-00042 Consolider
Project CPAN) and by GV (PROMETEUII/2013/007). SB acknowledges support from JAEPre programme (CSIC). GC acknowledges support from Marie Curie Actions (PIEF-GA-2011-298582).












\end{document}